\documentclass{aa}
\usepackage{graphicx}
\usepackage{natbib}

\usepackage{txfonts}

\usepackage{hyperref}

\newcommand{\dg}{$^{\circ}$} 
\newcommand{\rs}{$\mathrm{R_{\odot}}$} 
\newcommand{\bc}{\begin{center}} 
\newcommand{\ec}{\end{center}}  

\begin{document}

\title{Constraints on the variable nature of the slow solar wind with the Wide-Field Imager on board the Parker Solar Probe}

\author{S. Patsourakos \inst{1}
\and A. Vourlidas \inst{2}
\and A. Nindos \inst{1}
}
\institute{Physics Department, University of Ioannina, Ioannina GR-45110,
Greece\\
\email{spatsour@uoi.gr}
\and
The Johns Hopkins University Applied Physics Laboratory, Laurel, MD 20723, USA}
\authorrunning{S. Patsourakos et al.}
\titlerunning{Constraints on the variable nature of the slow solar with WISPR}

\date{Received  ; accepted }

 
  \abstract
  {The formation of the slow solar wind remains unclear as we lack a complete understanding of its  transient outflows.}
{In a previous work  we analysed the white-light coronal brightness as a function of elongation and time from Wide-Field Imager (WISPR) observations on board the Parker Solar Probe (PSP) mission  when PSP reached a minimum heliocentric distance of $\approx$ 28 \rs. We found 4-5 transient outflows per day over a narrow wedge in the PSP orbital plane, which is close to the solar equatorial plane.
However, the elongation versus
time map  (J-map) analysis supplied only lower limits on the number of released density structures due to the small spatial-scales of the transient outflows and line-of-sight integration effects. 
In this work we  place constraints on the properties of  slow solar wind transient mass release from the entire solar equatorial
plane.}
{We simulated the release and propagation of transient density structures in the solar equatorial plane for four scenarios: (1) periodic release in time and longitude with random speeds; (2) corotating release in longitude, periodic release in time with random speeds; (3) random release in longitude, periodic release in time and speed; and (4) random release in longitude, time, and speed. }
{The simulations were used in the construction of synthetic J-maps, which are  similar to the observed J-map. The scenarios with periodic spatial and temporal releases are consistent with the observations for periods spanning  3\degr-45\degr longitude and 1-25 hours. The four considered scenarios have  similar ranges (35-45 for the minimum values and 96-127 for the maximum values) of released density structures per day from the solar equatorial plane and consequently from the streamer belt, given its proximity to the solar equatorial plane during the WISPR observation.
 Our results also predict that density structures with sizes in the range 2-8 \rs\, covering 1-20 \%  of the perihelion could have been detectable by PSP in situ observations during that interval.}
{Our estimates of the release rates of density structures from the streamer belt represent a first major step towards assessing their contribution to the slow solar wind mass budget and their potential connection with 
in situ detections of density structures by PSP.}

\maketitle
\section{Introduction}

Only a few years after the theoretical prediction 
of the dynamic expansion of the solar corona into the heliosphere in the form of the solar wind by \citet{parker1958} was  its existence   
confirmed by the first in situ measurements in the interplanetary space \citep{gringauz1960,neugebauer1962}. At the most basic level, 
the solar wind at 1  au is classified into fast ($> ~$ 500 km/s) quasi-steady  and  slow ($< ~$ 500 km/s) variable  wind streams \citep[e.g.,][]{mccomas2000}. Although it is now clear that fast winds originate from polar coronal holes, the origins of the slow streams are more elusive \citep[see recent reviews][and references therein]{antiochos2012,abbo2016,cranmer17,vial2020}. The difference in variability between fast and slow streams seems to indicate differences in the source region and/or release  mechanism(s) behind these streams. Naturally, much research on slow wind focuses on how much it deviates from a steady release regime and what spatio-temporal scales and physical mechanisms are involved  \citep[e.g.][]{vial2020}. Recent observational studies by \citet{antonucci2023} ,\citet{baker2023}, and 
\citet{chitta2023} supplied
evidence in favour of the S-web slow-solar wind model \citep[][]{antiochos2011}. This model  postulates
that the  slow wind emanates from a network of narrow open-field corridors in the low corona that are connected to a web of separatrices and quasi-separatrix layers in the heliosphere.

\begin{figure}[h]
\centering
\includegraphics[width=0.5\textwidth]{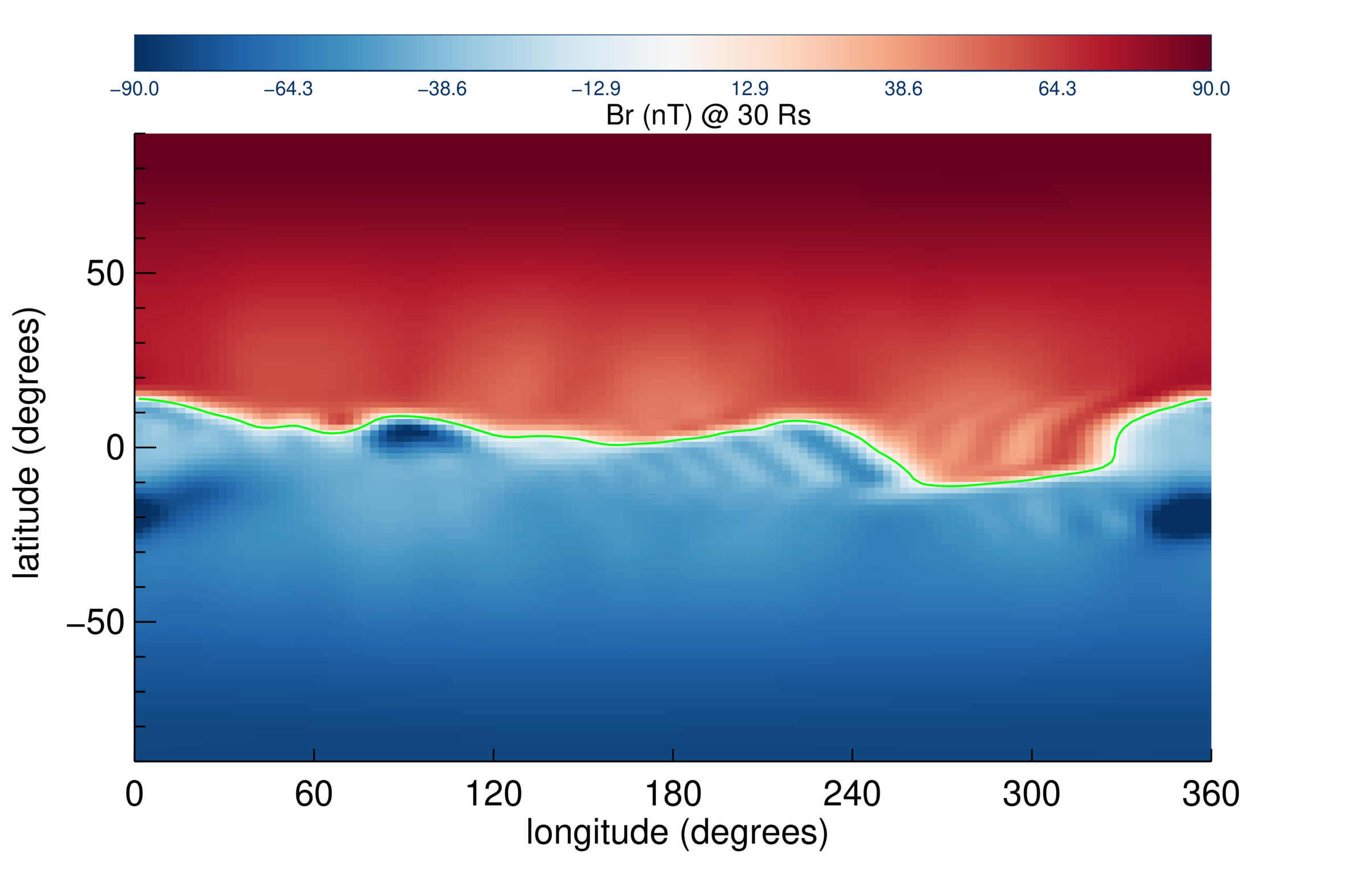}
\caption{Synoptic map of the  radial component of the heliospheric magnetic field (in nT) at 30 \rs, on 29 January 2020, 12 UT, from a medium-resolution run 
of the MAS code \citep[e.g.][]{mikic1999,riley2012} corresponding to Carrington
rotation 2226. The green line corresponds to the heliospheric neutral line. The simulation data are from \url{http://www.predsci.com/}.}
\label{fig:fpsi}
\end{figure}

The research is focused at the cradle of the solar wind, the solar corona. Coronagraphs and heliospheric imagers provide the main observational means for the study of the near-Sun solar wind. These instruments typically record the visible corona (K-corona) that results from the scattering of the photospheric radiation from the coronal free electrons. Because of the optically thin 
nature of this emission, the observed signals correspond to the line-of-sight (LoS) integral of the electron density. Therefore, the features seen in a coronagraph or a heliospheric image represent the overlap of density structures along the LoS.

The time series of images from the Large Angle and Spectroscopic Coronagraphs \citep[LASCO;][]{brueckner1995} on board the Solar and Heliospheric Observatory mission \citep[SOHO;][]{domingo1995} and from the Sun-Earth Connection Coronal and Heliospheric Investigation  \citep[SECCHI;][]{howard2008} instrument suite on board the twin spacecraft of the Solar TErrestrial RElations Observatory mission \citep[STEREO;][]{kaiser2008} have been providing ample evidence of intermittent density outflows in and around streamers. Some of the larger outflows have the shape of a blob, hence they are called streamer blobs \citep[][]{sheeley1997,wang1998}. Streamer blobs have been shown to be magnetic flux ropes resulting from intermittent magnetic reconnection between closed and open magnetic field lines \citep[e.g.][]{wang1998,sheeley2009,antiochos2011,higginson2018}. About two to six streamer blobs per day are recorded during solar minimum and maximum conditions, respectively, with typical speeds  of a few hundred km/s \citep[e.g.][]{wang1998,sheeley2009, roulliard2010}.  

The blobs, and more generally transient slow solar wind density structures, exhibit quasi-periodic behaviour   with periods ranging from  $\approx$ 1 hour   to 19 hours, and with radial sizes from $\approx$ 12 to 1 \rs\ or less \citep[e.g.][]{viall2010,viall2015,sanchez2017,stansby2018}. Magnetohydrodynamic (MHD) modelling  suggests that tearing mode instability may be a candidate mechanism as it can give rise to periodic mass release in streamers with periods from $\approx$ 20  hours  to 1-2 hours \citep{reville2020}.  On the other hand, it should be  remembered that these statistics are based on 1 au measurements that suffer considerable projection effects that tend to smear structure and confuse its origins. Specialized campaigns of high-cadence (5 min) and deep exposure (36 sec) SECCHI/COR2 observations have revealed higher spatio-temporal intermittency embedded in streamer flows \citep{deforest2018}. However, the observing distance and long LoS  limit  the amount of information that can be extracted from 1 au imaging observations. Outflows in coronagraphic and heliospheric imaging data are often analysed with J-maps \citep[][]{sheeley1997}. These are maps of elongation versus time of the brightness along a given position angle;  they offer an efficient method to assess coronal activity over long time periods.

\begin{figure*}[ht]
\centering
\includegraphics[width=0.6\textwidth]{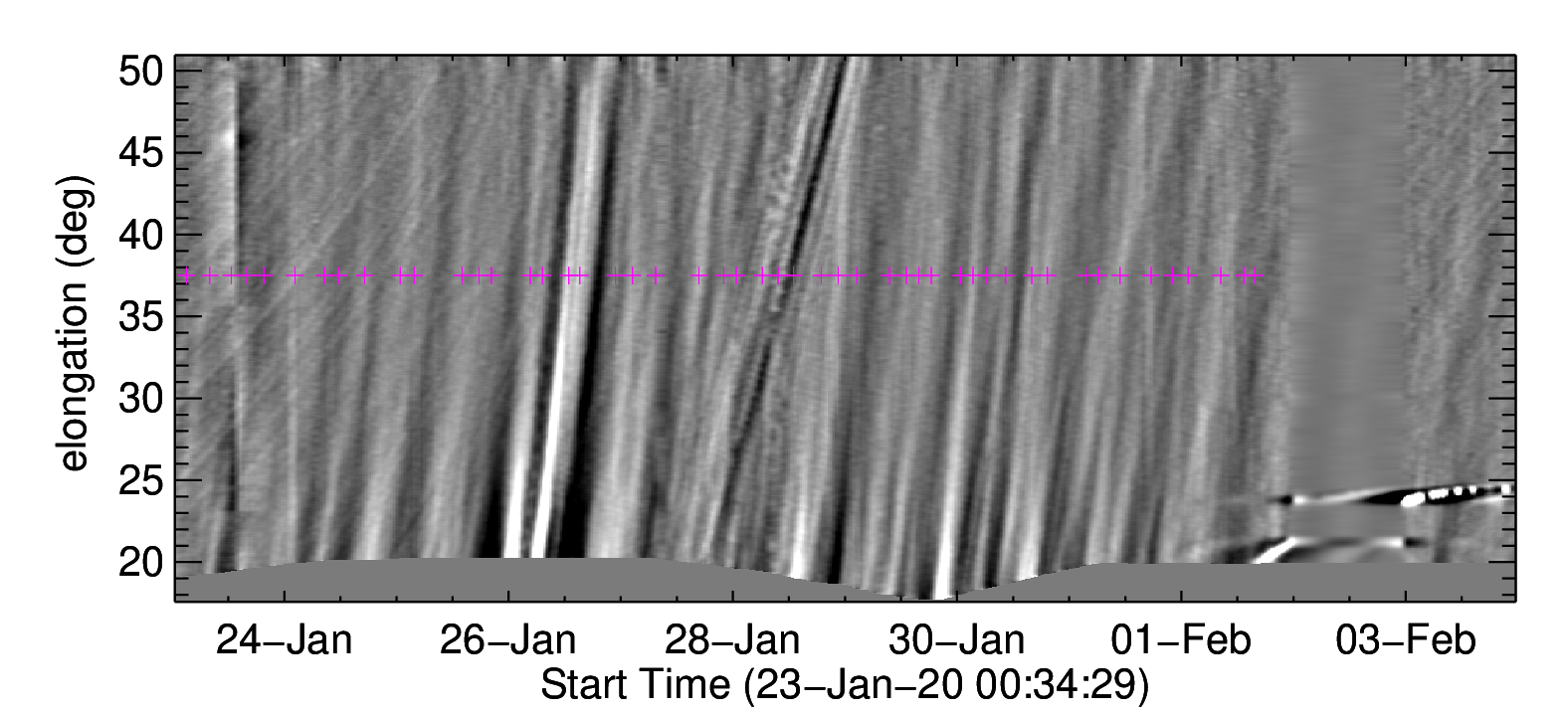}
\caption{WISPR-I J-map during PSP E4. The purple crosses correspond
to the automatically detected tracks. The figure is modified from Paper 1.} 
\label{fig:fpa1}
\end{figure*}

The Parker Solar Probe \citep[PSP;][]{fox2016} mission offers an unprecedented opportunity to break these constraints thanks to its unique design and payload. PSP's orbit brings it ever deeper into the corona, starting at at 35 \rs\ in 2018 and ending at 9.8 \rs in 2025, making it the first mission to measure and image the origins of the solar wind from within the corona. The  imaging payload consists of the Wide-field Imager  for Parker Solar Probe \citep[WISPR;][]{vourlidas2016} whose two telescopes have a combined radial field of view (FoV) of 13.5\degr-108\degr elongation from the Sun centre. Its coronal vantage point within the corona removes much of the LoS confusion intrinsic to 1 au imagers and allows unprecedented observations of the fine-scale structures of coronal mass ejections and background streamer structures \citep{Howard2019,rouillard2020a}. On the other hand, the rapidly changing viewpoint introduces a set of novel  challenges in the analysis of imaging data that do not pertain to the SOHO or STEREO observations from 1 au.

\citet[][hereafter Paper 1]{nindos2021} reported the first kinematic analysis of outflows in the WISPR images by developing a methodology to account for the rapidly varying viewpoint. They identified four to five quasi-linear tracks of outflowing features per day, during the fourth solar encounter of PSP (E4)
from 23 January 00:44:26  to 3 February, 2020 23:04:26 UT. The measurements were taken over a   one-degree wide wedge along the PSP orbit plane. The heliospheric neutral line  marking the streamer belt, during this period, was within 10\degr\ of the solar equatorial plane (Fig.  \ref{fig:fpsi}),
and hence it was  close to the PSP orbital plane (the PSP orbit is inclined by 3.4\degr\ relative to the equatorial plane). The detected outflows are therefore associated with streamers. Paper 1, however, did not investigate the implications of these results.

The core goal of our paper is to supply constraints on various properties of transient slow solar wind density structures, such as their spatio-temporal periods  and  total number. This is achieved via  Monte Carlo  simulations of synthetic WISPR-I J-maps, spanning E4, and  for different scenarios, encapsulating plausible properties of density structure release motivated by pertinent observations and modelling, which are then juxtaposed with the observed J-map in  Paper 1. Our paper has the following outline. We describe the simulations in Sect. 2 and the construction of the synthetic J-maps in Sect. 3. We present the results in Sect. 4 and discuss them in Sect. 5. We conclude in Sect. 6.

\section{Simulations of transient solar wind density structures} 

\subsection{The need for simulations}

In J-maps, any outward propagating  density structure will create a track with  either an upward  (accelerating) or a downward (decelerating) curve (see Fig.~\ref{fig:fpa1}). The slope is a function of the feature's own kinematics and of projection effects. In other words, a given feature's kinematic profile cannot be taken at face value, but requires analysis and the use of various constraints, as has been described in numerous publications \citep[e.g.][]{sheeley1997,davies2009,rouillard2011,lugaz2009,liu2010,sheeley2010}. Depending on the particulars, the analysis results in speed, direction, and solar origin estimates. 

Projection effects are always a concern in the analysis of coronal observations, but are especially important in the PSP case due to the rapidly varying heliocentric distance. A key property of Thomson scattering is that the location of maximum scattering efficiency lies on the so-called Thomson surface (TS) \citep[][]{vourlidas2006}, which is a sphere with a diameter equal to the Sun--observer distance.  
The TS dependence on the Sun--observer distance implies that the recorded intensity in inner heliospheric observations is more heavily weighted towards the near-spacecraft environment than a similar observation from 1 au. In the case of WISPR, this effect turns the instrument into a local imager \citep{vourlidas2016};  however, it has a rapidly varying sensitivity due to the orbit. The temporal variation of both the LoS integration and the distance between observer and feature will affect the shape of the tracks in J-maps to various degrees, and therefore will influence the interpretation.
In addition,
the small size of transient outflows and the LoS
effects discussed result in lower limits for the actual numbers of released density structures from the streamer belt.
Therefore, we construct synthetic observations under a variety of  scenarios to further explore the potential of WISPR-based analysis.

\subsection{Constructing simulated J-maps}

With our simple simulations we  emulate transient releases of slow solar wind parcels.   
The simulations consist of idealized density structures released from the solar equatorial plane during the  PSP fourth solar encounter (23 January to 3 February 2020, $\approx$ 11.9 days). We consider four scenarios to encapsulate a wide range of plausible properties of density structure release. The scenarios are motivated by the observations and modelling discussed in the Introduction, and are described below. The structures are released at a heliocentric distance of 5 \rs\ for all four scenarios \citep[e.g.][]{sheeley1999}. They propagate radially at a constant speed randomly drawn from the interval [100,400] $\mathrm{km/s}$, consistent with the observations of  streamer blobs. 
\begin{itemize}
    \item \textbf{Scenario 1, periodic release longitudes and release times and random speeds:}
    In Scenario 1 $\mathrm{n_{theta}}$ density structures are released every $dt$ minutes from equally spaced longitudes in the equatorial plane. Hence, a total of $N_{tot}=\frac{n_{theta} {T}_{E4}}{dt}$  structures are released during the duration $T_{E4}$ of E4. We consider nine $dt$ values, equally spaced logarithmically in the interval [1, 25.6] hours, and six $n_{theta}$ values, equally spaced logarithmically in the interval [4, 128]. This scenario corresponds
to density structures released every  $\approx$ 3 to 90 degrees across the solar equatorial plane.
    \item\textbf{Scenario 2,  corotating release longitudes, periodic release times and random speeds:} In Scenario 2 $n_{theta}$ equatorial density structures are released every $dt$ minutes during E4, from equally spaced Carrington (i.e. corotating) longitudes. This scenario corresponds to fixed release sites on a rotating Sun.     
    \item\textbf{Scenario 3,  random release longitudes and speeds, periodic release times:} In Scenario 3 $n_{theta}$ density structures are released every $dt$ minutes from randomly selected equatorial longitudes. We note that here    the $n_{theta}$ and $dt$   grids and total number of released density structures are the same  in Scenarios 1--3.
    \item\textbf{Scenario 4, random release longitudes, release times, and speeds:} In scenario 4 $N_{tot}$ density structures are released at randomly distributed times and from randomly distributed equatorial longitudes. $N_{tot}$ takes 13 values equally distributed logarithmically from the interval [20,81920]    and was chosen in order to encompass the corresponding intervals employed in Scenarios 1--3. As for Scenarios 1--3 a logarithmic grid was used. 
\end{itemize}

The employed release times are consistent with  observations and
with the modelling of quasi-periodic mass release discussed in the Introduction, and they span
the interval from 1 hour   to 20 hours \citep[e.g.][]{viall2010,viall2015,sanchez2017,stansby2018,reville2020}. Likewise, for the spatial scales of release (i.e. $n_{theta}$) synoptic maps
of the coronal brightness in the outer corona using rotational tomography constructed by \citet[][]{morgan2020}  showed
that in some cases streamers repeat themselves over several tens of degrees. Streamers, as discussed in the Introduction, represent potential sites
of transient slow solar wind release.
Moreover, the observations reported by
\citet{sanchez2017} suggest a 15-degree separation of transient slow solar wind outflows, as captured by
heliospheric imagers.
Finally, for transient mass releases associated with coronal jets from the quiet Sun, the relevant spatial scale
here    \citep[e.g.][]{nour2016,nour2023} is the chromospheric network size, which corresponds to $\approx$ 2.5 
degrees. 

As both the employed $n_{theta}$ and $dt$ studied intervals used in Scenarios 1--3 span   an order of magnitude, and to
save on computing time given the large number of simulated structures and generally the large number of iterations   inherent to Monte Carlo methods, we populated these intervals 
with logarithmic grids. The same applies to the choice of a logarithmic grid for  
$N_{tot}$ for Scenario 4. For this scenario the range of $N_{tot}$ was chosen to
be consistent with the $N_{tot}$ range used in Scenarios 1--3. The number 
of points in each considered interval was chosen in order to cover the corresponding
interval with a reasonable number of points.

Using finer grids of the input parameters $dt$ and $n_{theta}$ would lead to   higher precision in the determination of  the ranges of these parameters that are consistent with the observations.  However, this is beyond the scope of the present study;   we do not wish to fine-tune $dt$ and $n_{theta}$
to the observations, but rather to supply some estimates of the ranges of these parameters that are consistent with the observations.

For simplicity and computational efficiency (i.e. to avoid LoS integrations over a large number of discrete structures)
\citep[e.g.][]{patsourakos1997,liewer2019,nistico2020}, 
we treated the density structures as zero-mass and point-like entities. Therefore, the implicit assumption in the construction of the synthetic J-maps is that the density structures are sufficiently massive to be detected by WISPR. We also did not take into account the 3\degr\ offset between the solar equatorial plane ecliptic and the PSP orbital plane because visible light structures are generally wider than 3\degr.
 Moreover,
this small offset does not impact the calculation of elongations
of structures released close to the PSP orbital plane 
\citep{liewer2020}. 
 \begin{figure}[!h]
\centering
\includegraphics[scale=0.1]{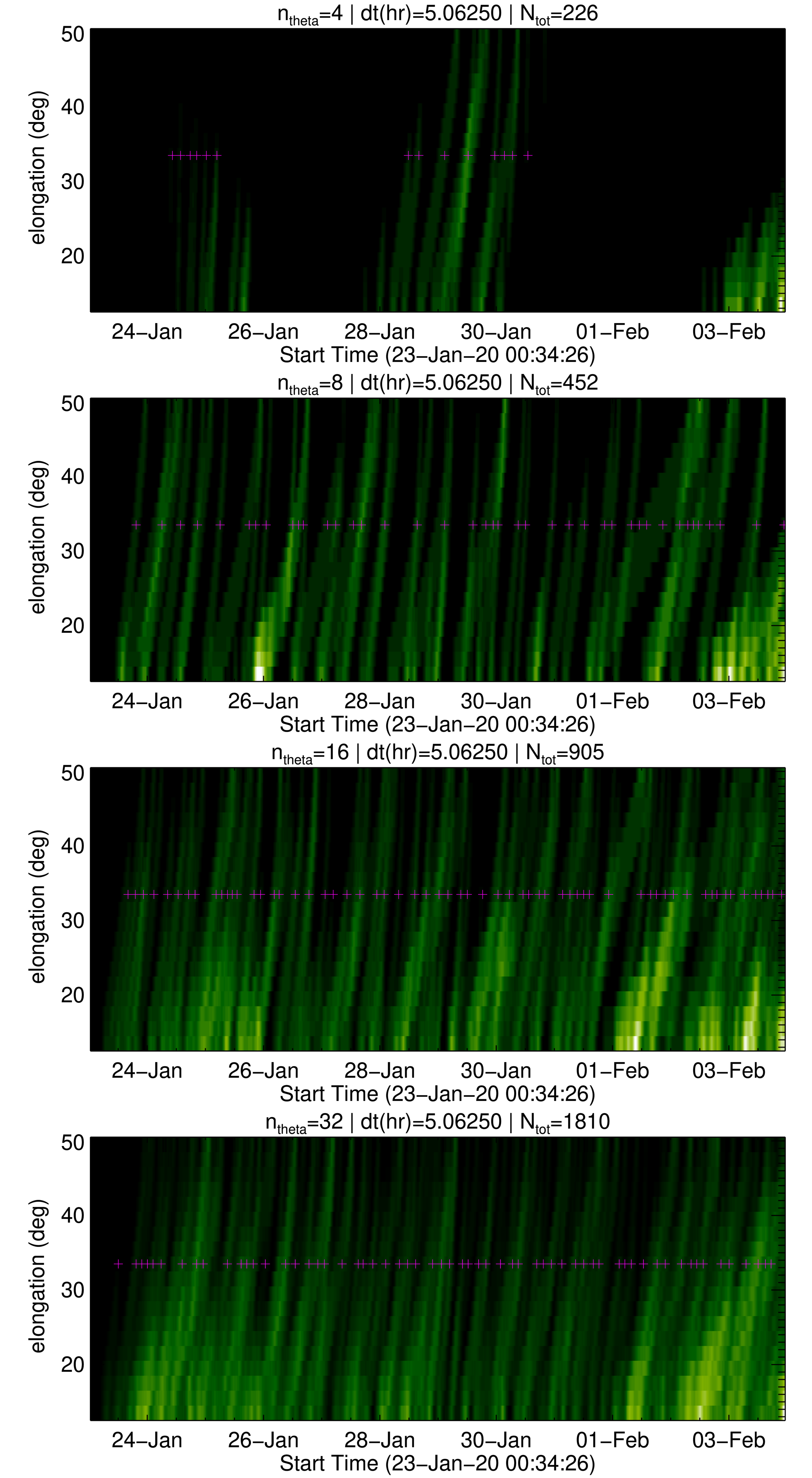}
\caption{Example synthetic J-maps for Scenario~1. The colour-coding from black to green to white corresponds to increasing J-map values. The purple crosses indicate the detected tracks.}
\label{fig:fexpamps1}
\end{figure}

\begin{figure}[!h]
\centering
\includegraphics[scale=0.1]{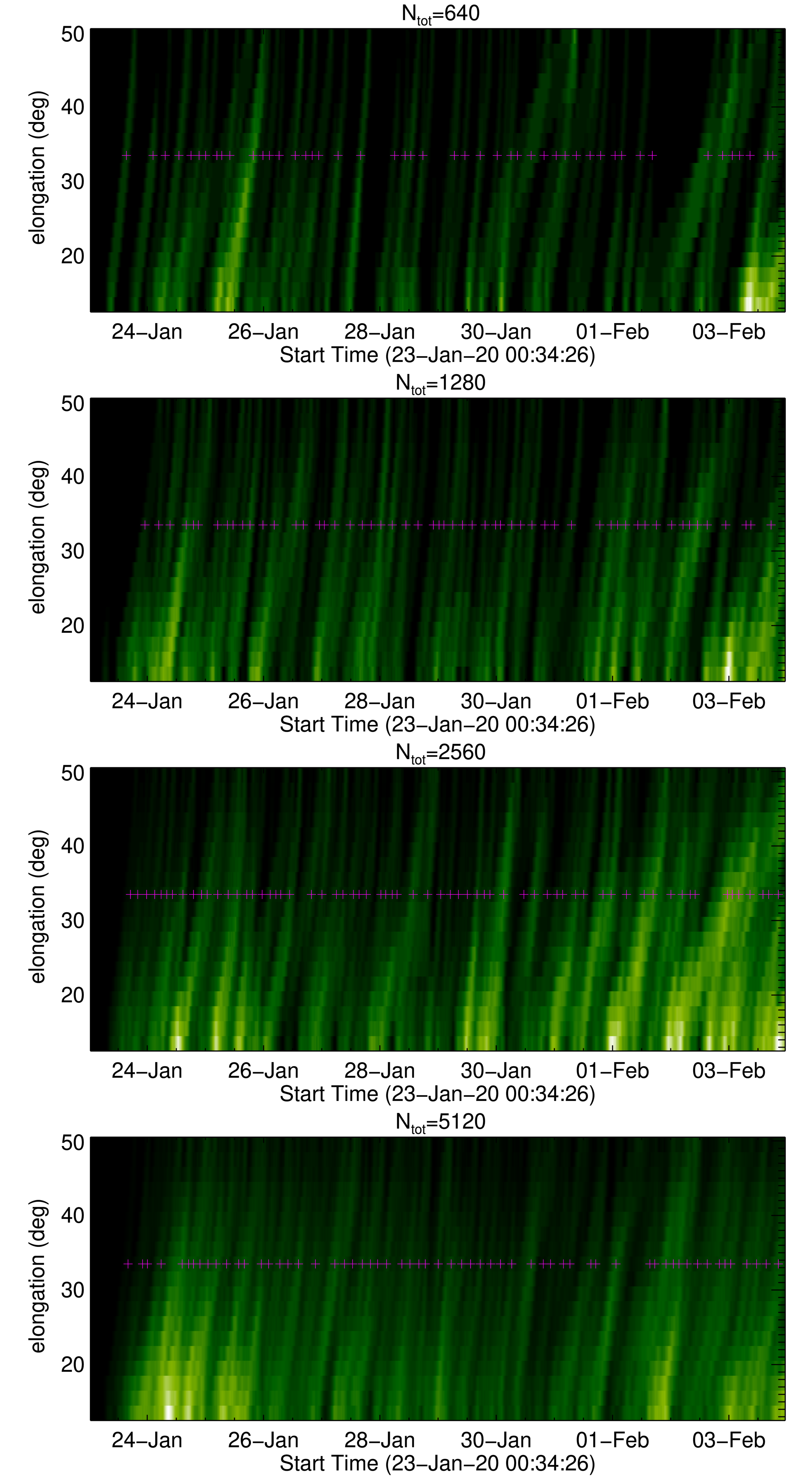}
\caption{Same as Fig.~\ref{fig:fexpamps1}, but for Scenario 4.}
\label{fig:fexamps4}
\end{figure}
Next, we performed 1000 Monte Carlo simulations for each scenario resulting in a large number of simulations and J-maps. Namely, we obtained 48,000 synthetic J-maps for Scenarios 1--3 and 13,000 for Scenario 4. For each Monte Carlo simulation
we drew randomly selected parameter(s) as needed for each scenario's specifics.

\begin{figure}[!h]
\centering
\includegraphics[width=0.45\textwidth]{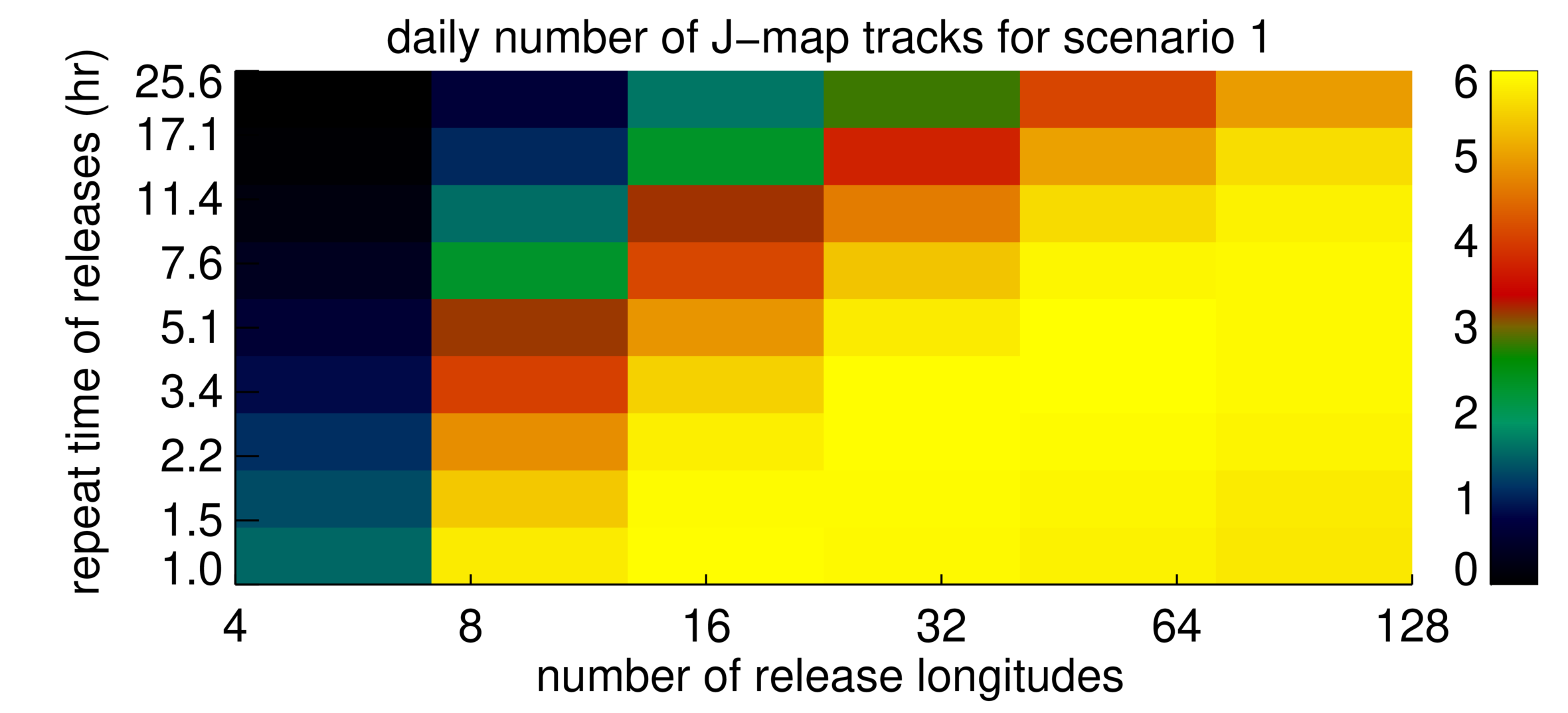}
\includegraphics[width=0.45\textwidth]{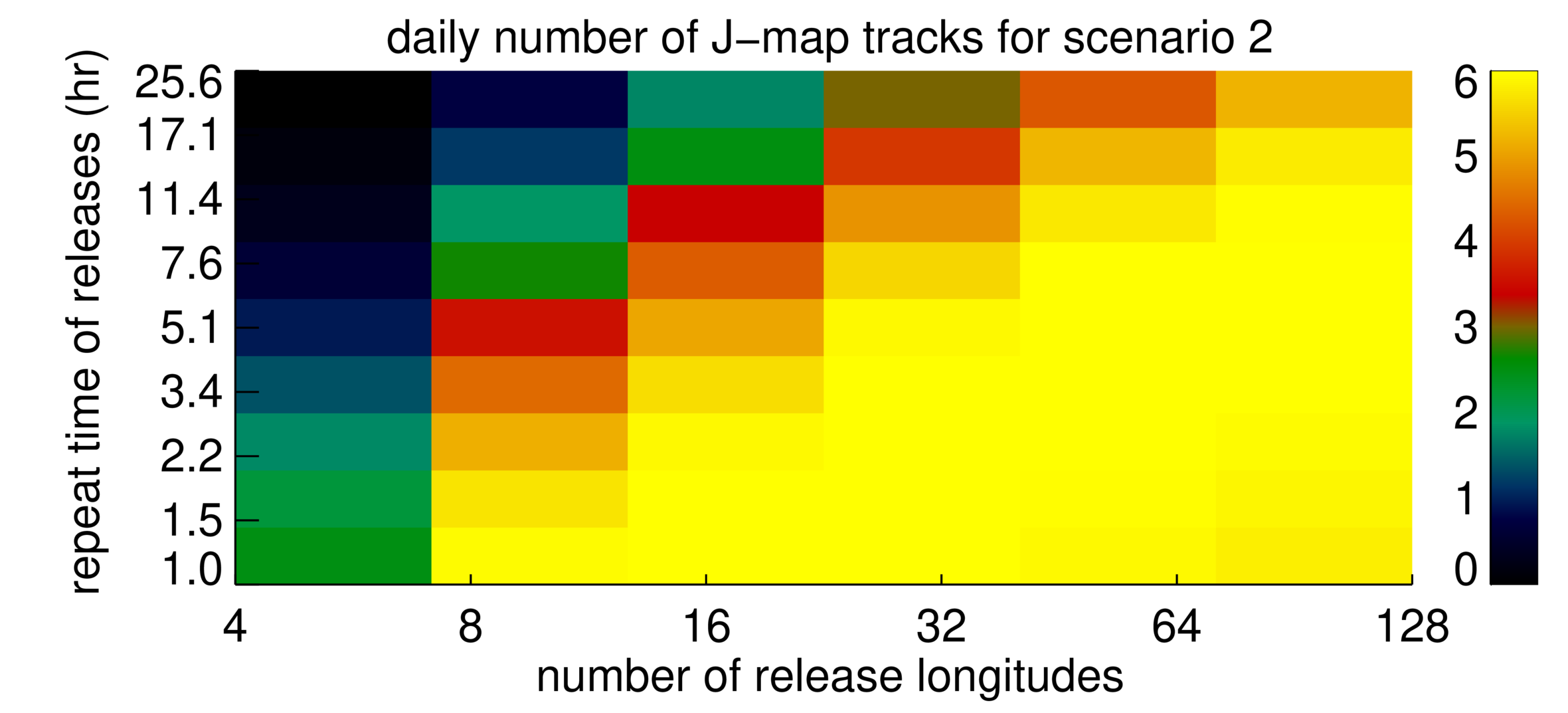}
\includegraphics[width=0.45\textwidth]{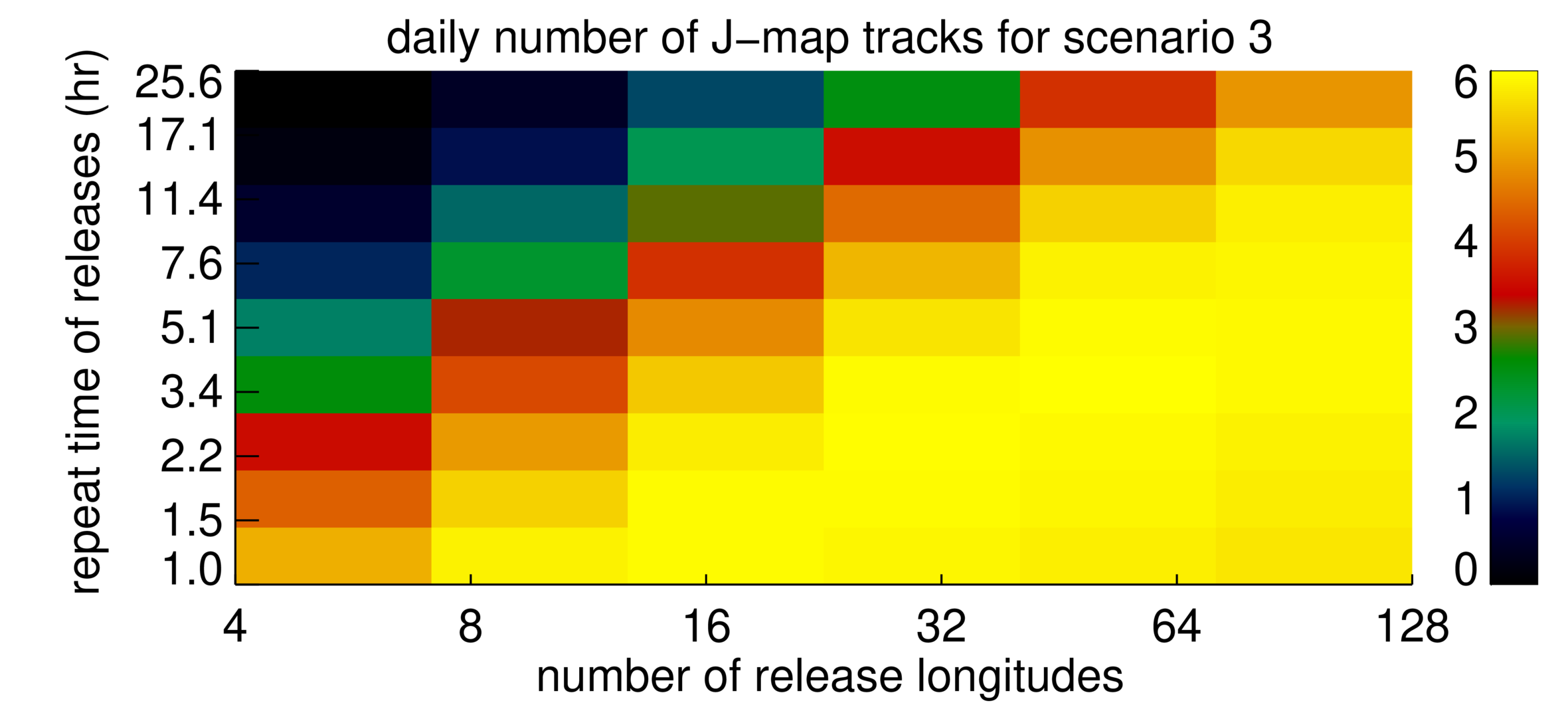}
\caption{Average daily number of detected tracks ($N_{tracks}$) in   synthetic J-maps from 1000 Monte Carlo simulations for Scenarios 1 (top), 2 (middle), and 3 (bottom) as a function of the number of release longitudes ($n_{theta}$) and repeat time ($dt$)  of the corresponding density structures.}
\label{fig:ftracks1}
\end{figure}
\section{Construction of synthetic J-maps}

We constructed J-maps for every Monte Carlo simulation as follows. Using a 20 min cadence to emulate the WISRP-I synoptic programme cadence, we recorded a density structure if the following two conditions were met: (1) its elongation lies within the elongation range of the WISRP-I FoV (that is 13.5-50\degr); (2) the density structure lies inside the corresponding TS. We note here that the vantage point of each considered timestamp during E4 in our simulations followed the actual motion of the PSP spacecraft with respect to the Sun during that interval; in other words,  each vantage point, and hence corresponding FoV, was determined from the
corresponding PSP location.

Next, for each  timestamp in our simulations, we counted the number of  density structures satisfying the two detection criteria over an elongation grid with a bin size of 2 degrees spanning the WISPR-I FoV, resulting essentially
to a histogram of detections versus elongation.
Time-stacking these histograms resulted into the simulated J-maps,
which obviously do not record brightness, as the actual J-maps do. However, non-zero J-map values at any given time-elongation pair imply the presence of  density structure(s), 
and therefore record the outflows. Large (small) J-map values correspond to a large (small) number of density structures for a given elongation-time pair. We applied a box-car filter to smooth local variations in the synthetic  J-maps.  J-map examples for Scenarios 1 and 4 are shown in Fig.~\ref{fig:fexpamps1} and \ref{fig:fexamps4}, respectively. Similar maps were obtained for the other two scenarios, but are not presented here for the sake of brevity. The lack of tracks at the beginning of the synthetic J-maps is due to the finite start time and reflects the time it takes for the first density structures to enter the WISRP-I FoV. The synthetic J-maps exhibit  quasi-linear tracks similar to those in the observed J-map of Fig. \ref{fig:fpa1}. In addition, the track duration in
the simulated J-maps (i.e.  the time
it takes the associated density structure to traverse the WISPR-I FoV) is 
of the order of one day, similar to what was found for the
observed J-map in Paper I. In addition, we can observe tracks in the simulated J-maps  in Figs. ~\ref{fig:fexpamps1} and \ref{fig:fexamps4}, which become progressively vertical. This  
indicates outflows approaching PSP \citep{liewer2019}. 
As expected, the track density follows the number of simulated structures (check the value of $N_{tot}$ in the captions of Figs. \ref{fig:fexpamps1} and \ref{fig:fexamps4}). 

\begin{figure*}[!h]
\centering
\includegraphics[width=0.5\textwidth]{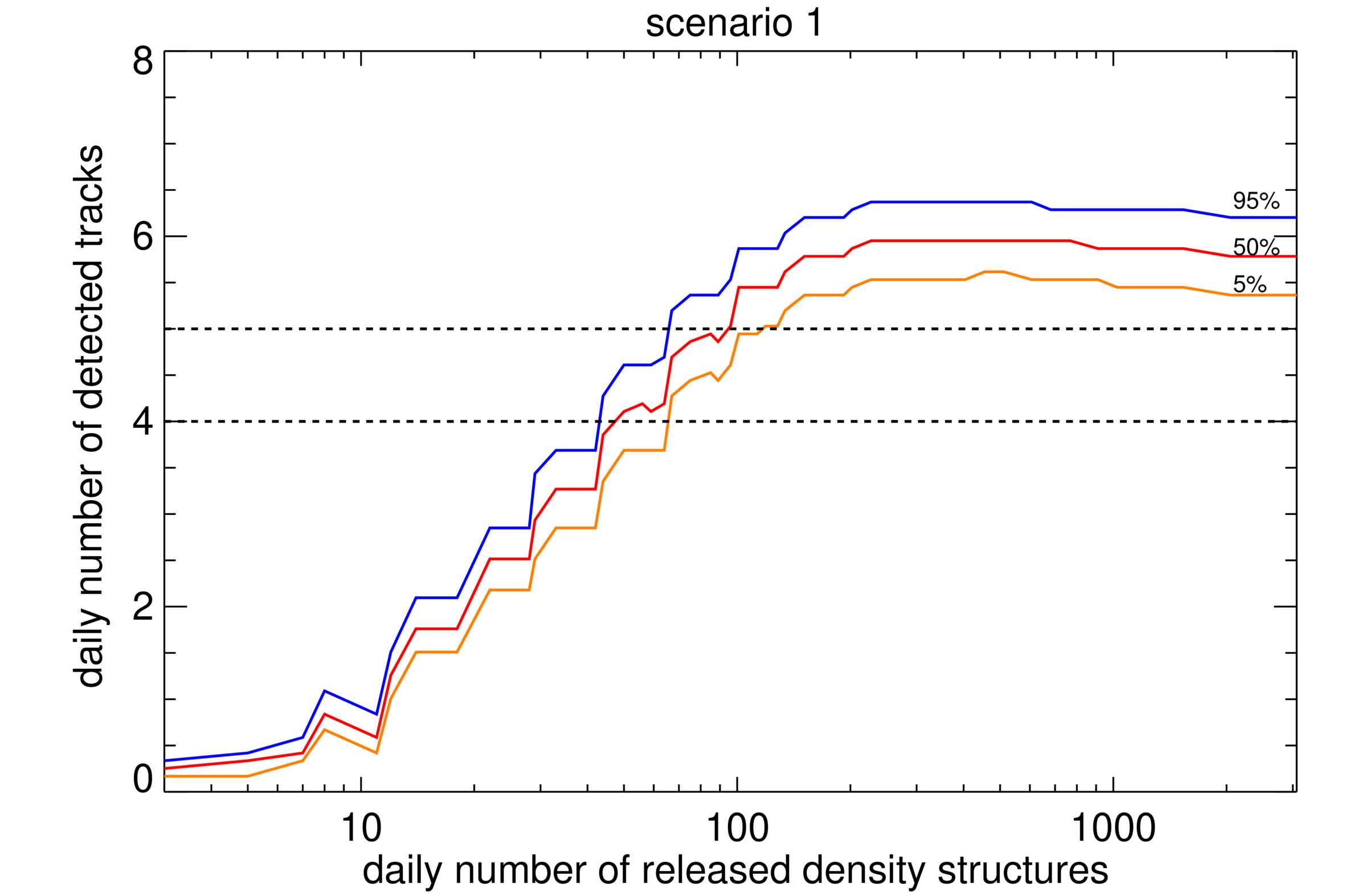}\includegraphics[width=0.5\textwidth]{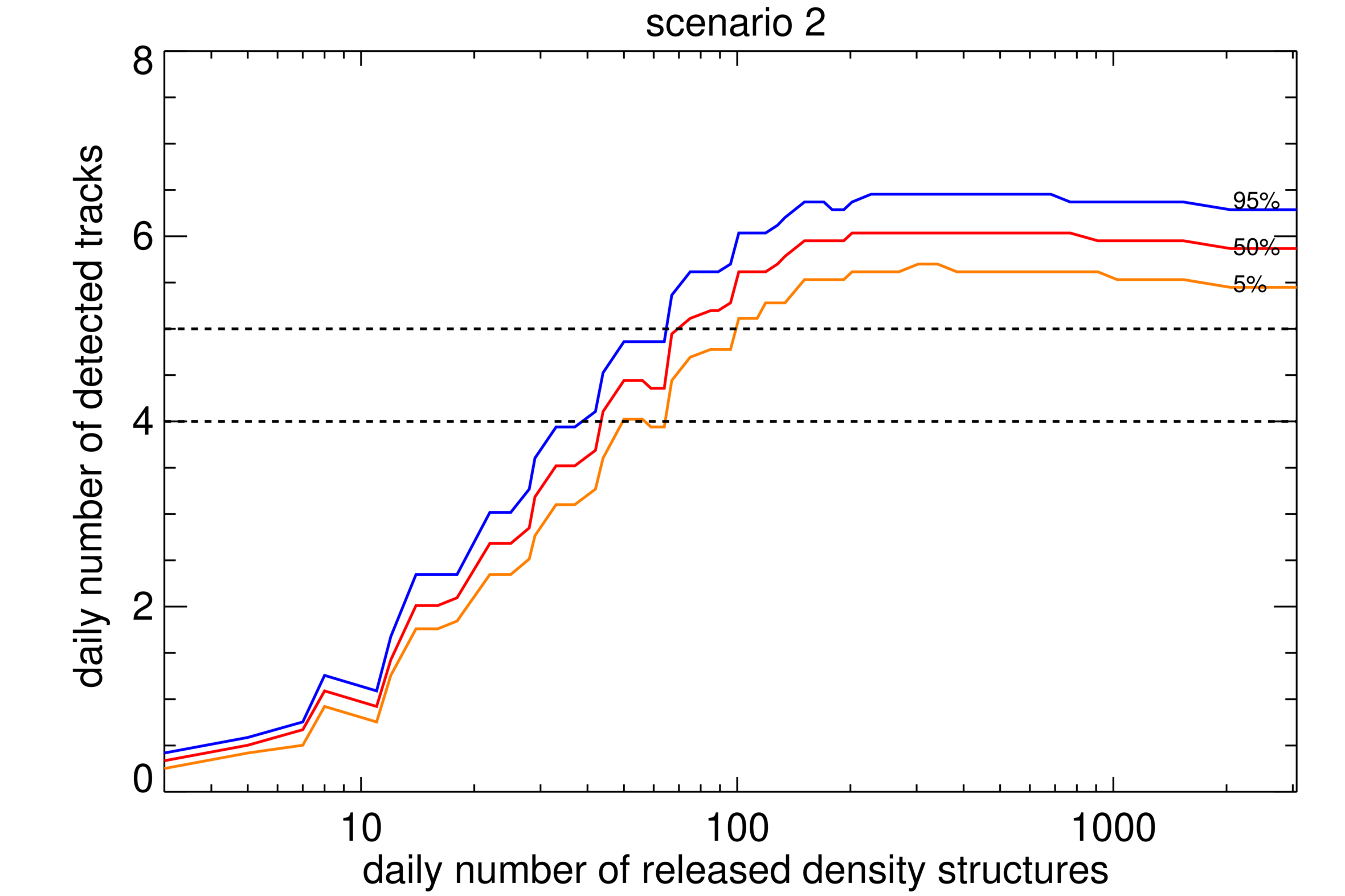}
\includegraphics[width=0.5\textwidth]{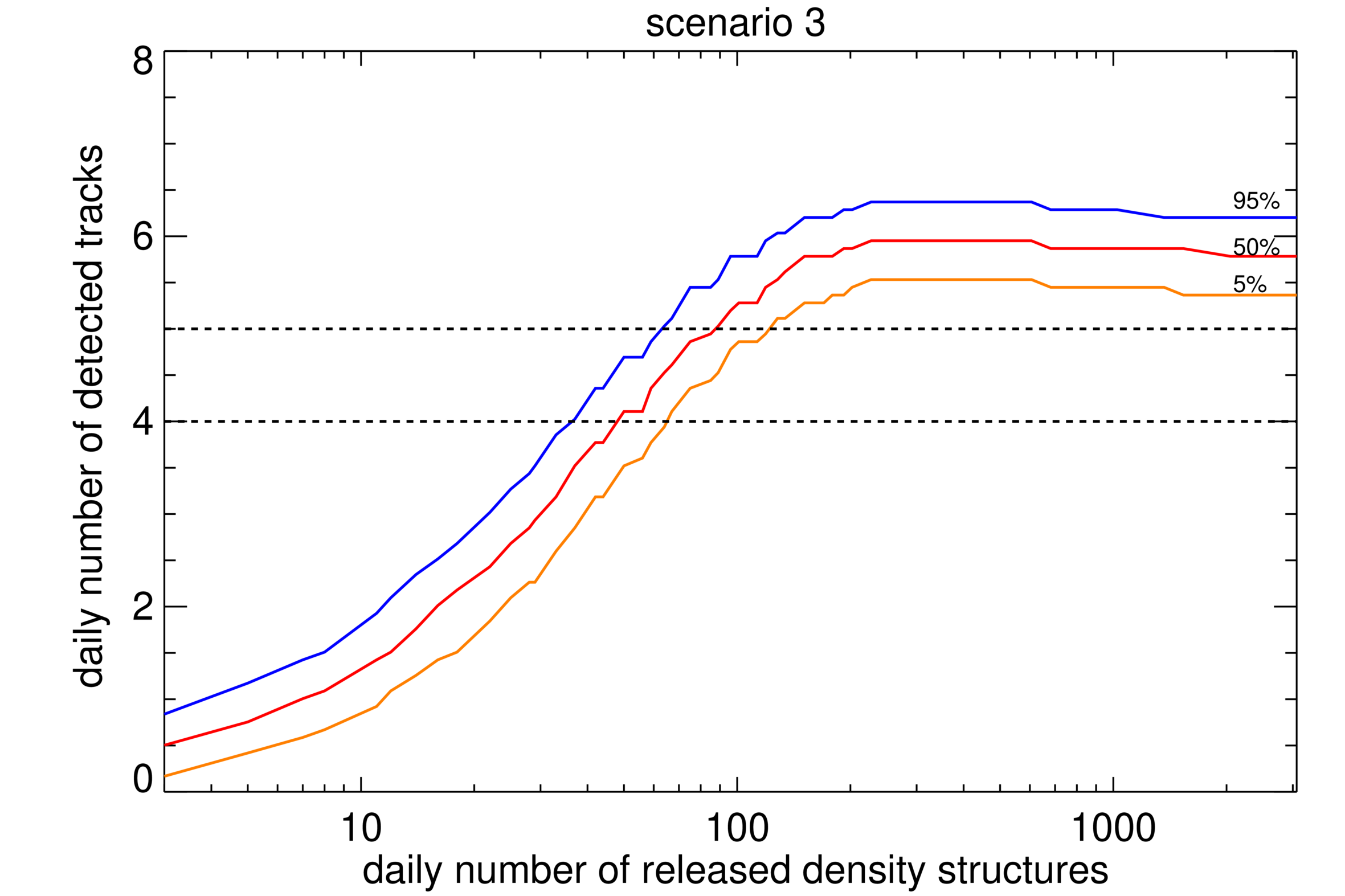}\includegraphics[width=0.5\textwidth]{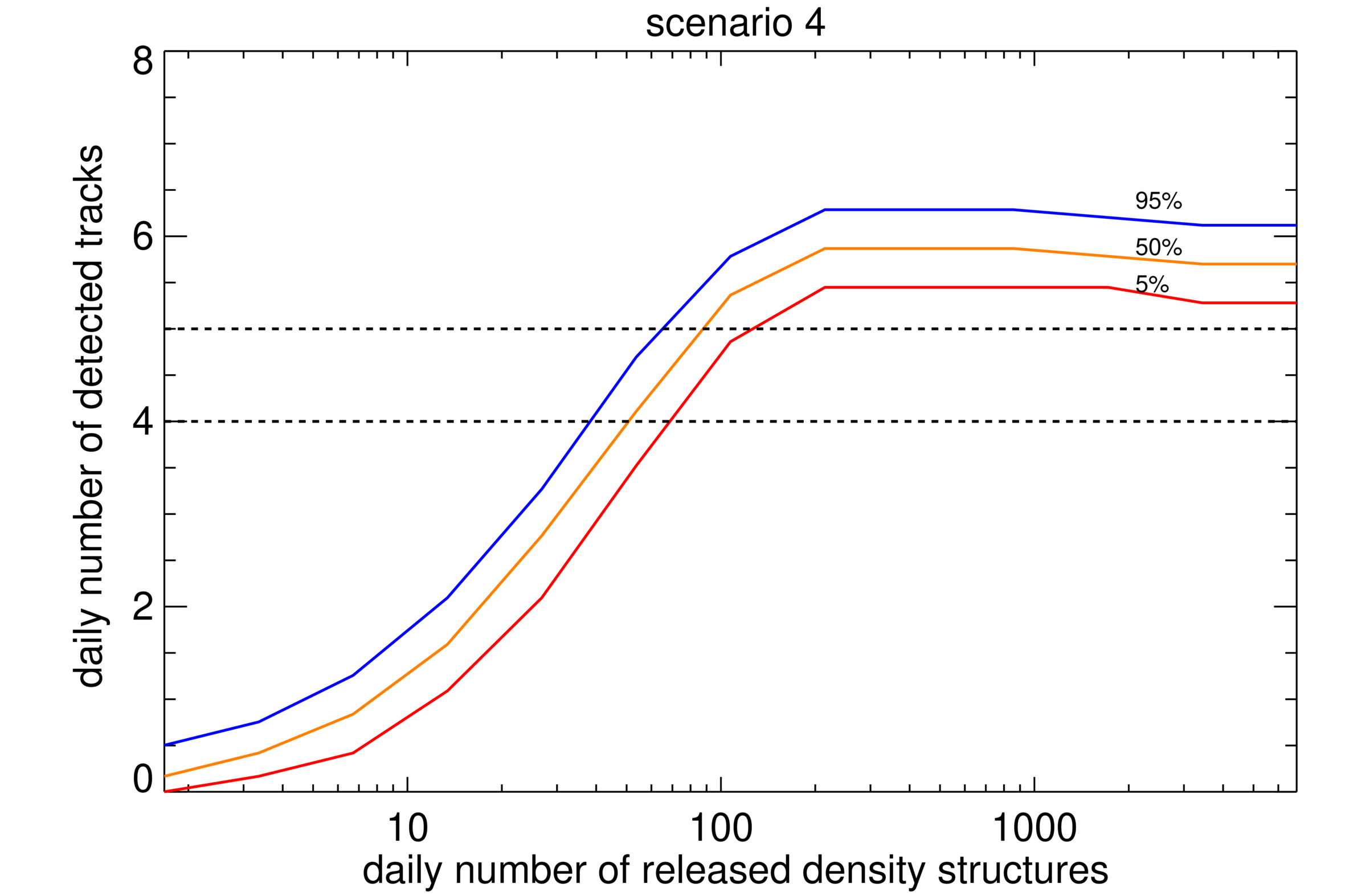}
\caption{Quartiles 
of the daily number of detected tracks for the 1000 Monte Carlo simulations (5\%\ in orange, 50\% in red line, and 95\% in blue) for Scenarios 1 (top left), 2 (top right), 3 (bottom left), and 4 (bottom right).  The two horizontal dashed lines correspond to four and five daily J-map tracks, consistent with the Paper 1 observations.}
\label{fig:fs1}
\end{figure*}

Given the large number of our simulations, we employ a simple method to automatically detect tracks in the synthetic J-maps. The method searches for local maxima in J-map time series created by averaging  the J-map over a 4\degr \ wide elongation bin centred at the middle of the WISRP-I FoV, which ensures the inclusion of transient outflows
that traversed a significant part of the relevant FoV.
The detected  local maxima 
is consistent with the temporal width (i.e. the duration)
of the tracks of the J-map in Paper 1, which 
resulted from the calculation of the autocorrelation function of the J-map time series at the centre of WISPR-I FOV of the J-map in Paper 1.

We validate the track-detection algorithm on the observed J-map in Paper 1
(see Fig.~\ref{fig:fpa1})
and the sample simulated J-maps (Figs. 3 and 4). The purple crosses in Figs.~\ref{fig:fpa1}--\ref{fig:fexamps4} give the locations of the detected tracks, which are largely consistent with visual inspection.

\section{Results}

To derive constraints on the  properties of the density outflows, 
we first examine the dependence of the number of detected tracks on the spatio-temporal periods of the released
density structures. Figure~\ref{fig:ftracks1} is a colour representation of the daily average (over the 1000 Monte Carlo simulations per scenario) of detected tracks per day ($N_{tracks}$) as a function of release longitude ($n_{theta}$) and time ($dt$) for Scenarios 1--3.

In addition, the higher the number of released density structures,   
the larger the $N_{tracks}$. 
A large portion of the plots corresponds to $N_{tracks}$ equal to or greater than six (corresponding to yellow). We   come back to this later on.
Periods of transient density structure release are given in
Table \ref{tab:table1}. In  Fig.~\ref{fig:ftracks1} and
Table \ref{tab:table1} we note that the  rather extended ranges
of both the spatial ($\approx$ 3\degr\ to 45\degr)
and temporal ($\approx$ 1-25 hours) release periods are 
consistent with the observations.
The temporal periods of the density structures discussed above are consistent with  observational \citep[e.g.][]{viall2010,viall2015,sanchez2017} or theoretical \citep[e.g.][]{reville2020} studies of periodic mass release from streamers.
\begin{figure}[ht]
\centering
\includegraphics[width=0.5\textwidth]{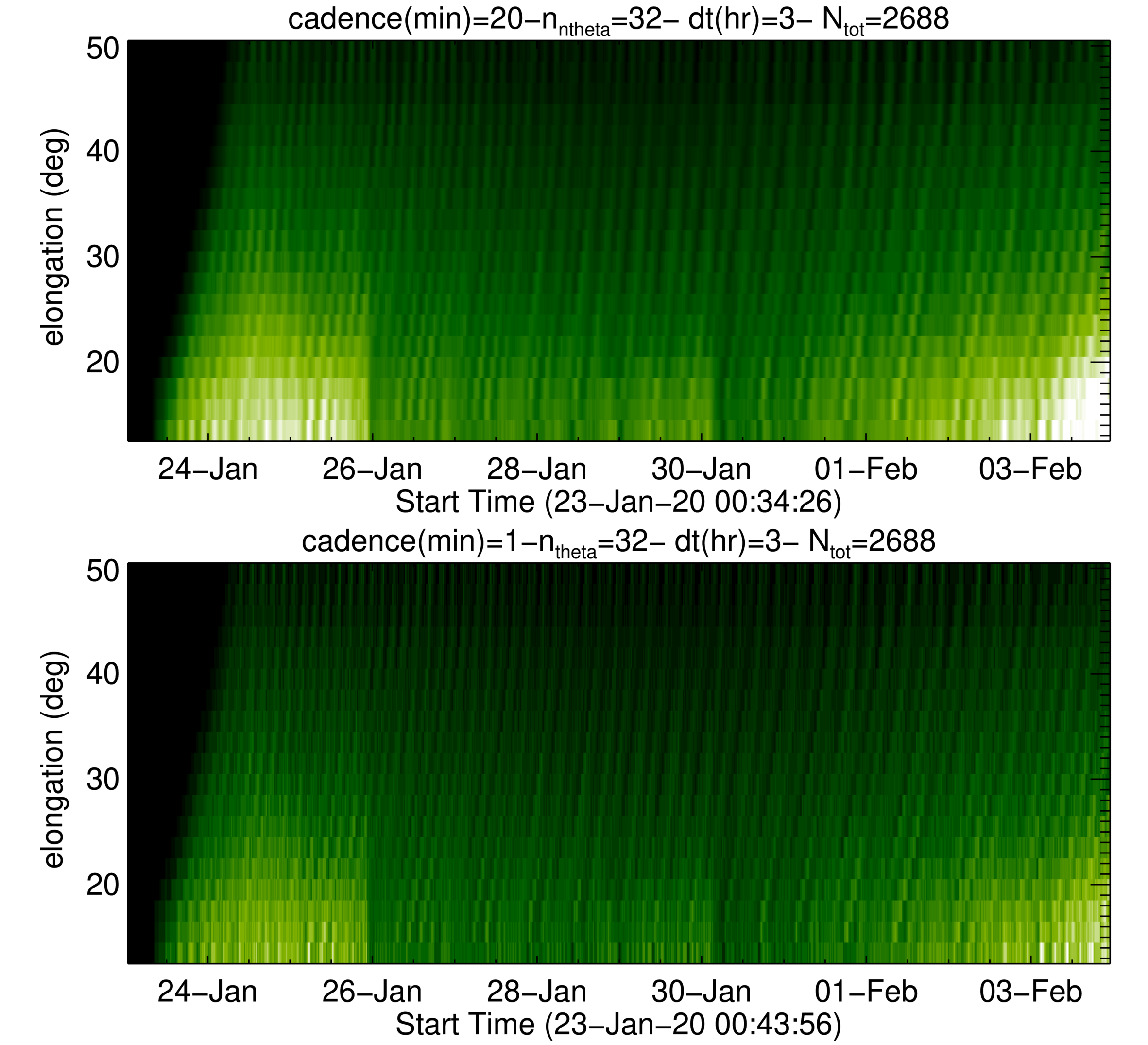}
\caption{Synthetic J-maps resulting from Scenario 1     considering the same speed (200 km/s) for all density structures. The only difference between the two J-maps is the cadence: 20 min (upper panel) and 1 minute (lower panel).}
\label{fig:fcad}
\end{figure}

\begin{table*}[htp]
\begin{center}
\caption{Spatio-temporal release periods of
density structures for Scenarios 1--3 leading to
four or five detected tracks in the synthetic J-maps per day, as resulting from the observed J-map.}
\begin{tabular}{ccc}
\hline 
\textbf{scenario} & \textbf{spatial period (\dg)} & \textbf{temporal period (hours)} \\
1: periodic release longitudes and release times and  random speeds  & 3-45 & 1.5-25 \\
2: corotating release longitudes, periodic release times and random speeds & 3-45 & 1.5-25 \\
3: random release longitudes, periodic release times and random speeds & 3-45 & 1.0-25\\ 
\hline
\label{tab:table1}
\end{tabular}
\end{center}
\end{table*}

\begin{figure}[h]
\centering
\includegraphics[width=0.5\textwidth]{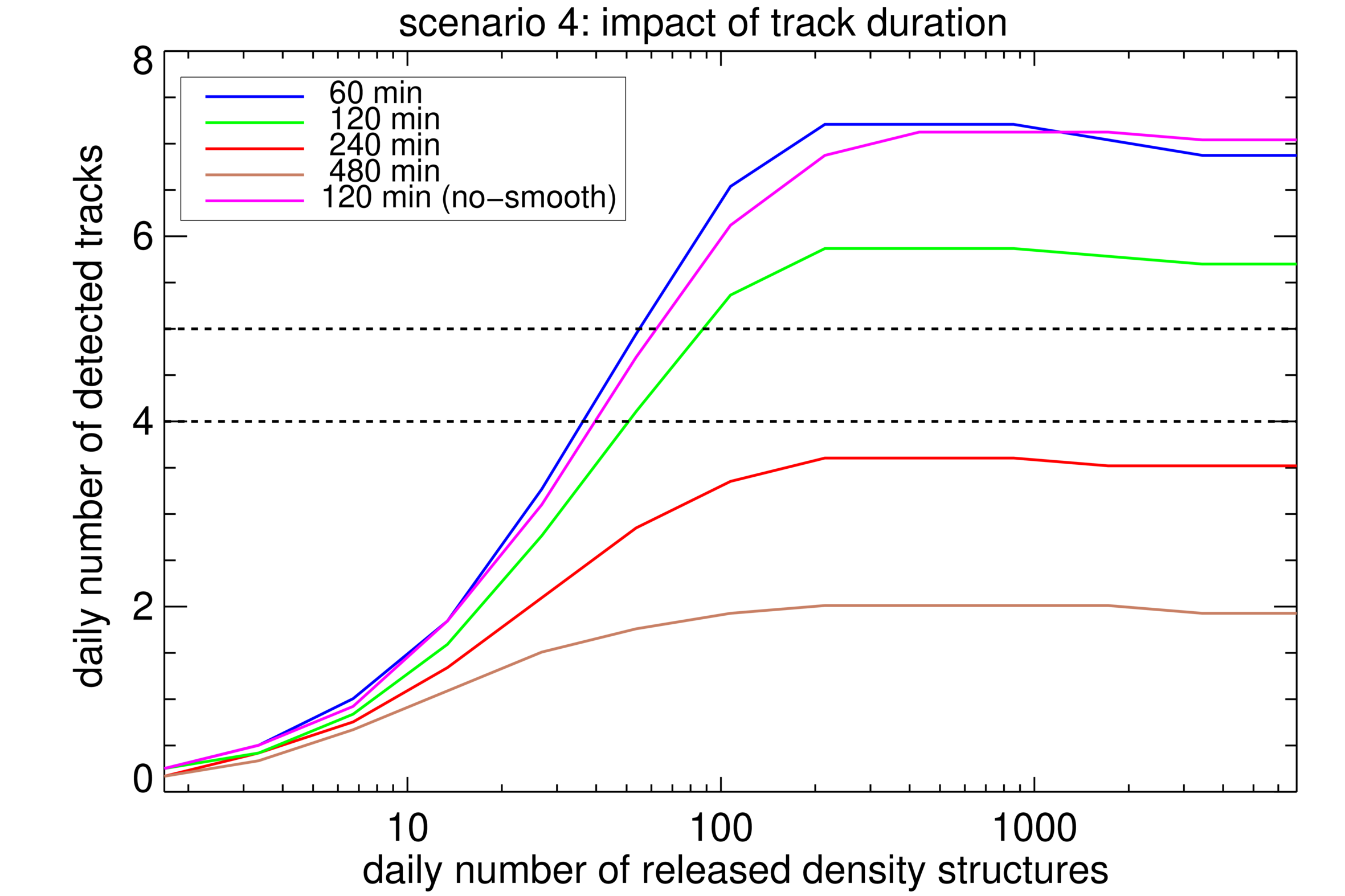}
\caption{50 \% quartiles of the daily number of detected tracks for the 1000 Monte Carlo simulations for Scenario 4 for J-map track duration (i.e. the interval between detected tracks; see Sect. 3) of 60, 120, 240, and 480 min (blue, green, red, brown lines, respectively). The purple line corresponds to results to when no smoothing is applied to the J-maps and the same temporal width as in our original study is used. The two horizontal dashed lines correspond to four and five daily J-map tracks, consistent with the observed J-map in Paper 1.}
\label{fig:fsen}
\end{figure}

Next, we examine the daily dependence of the number of detected tracks
versus the number of released structures for all four scenarios (Fig.~\ref{fig:fs1}). 
We note that the x-axis values for Scenario~4 strictly correspond to the averaged daily number of released density structures. In Fig. \ref{fig:fs1}  we display the daily 5, 50, and 95\% percentiles of the number of detected tracks  distributions for the corresponding 1000 Monte Carlo simulations as a function of the daily number of released structures. It should be noted in Fig.~\ref{fig:fs1}  that the abscissae
of the corresponding panels, containing the daily number of detected tracks 
resulting from the synthetic J-maps, are not expressed in integer values; the same applies to Fig.~\ref{fig:fsen}. This choice
is dictated by our main goal in building these figures. We decided  to use
the  $N_{tot}$-$N_{tracks}$ curves resulting from the simulations to infer 
$N_{tot}$, which is consistent with the results of Paper 1. This is achieved by essentially finding the intersection between the   $N_{tot}$-$N_{tracks}$ curves and horizontal lines at $N_{tracks}$ equal to 4 and 5, which is consistent with the 
observations in Paper 1. Rounding the $N_{tracks}$ values would obviously decrease the accuracy at which $N_{tot}$ is calculated.
All scenarios show a similar behaviour.  The number of detected tracks rises slowly for small numbers ($<10$) of released structures;  it increases rapidly with increasing number of structures before reaching a plateau of six detections  beyond about 100-200 released structures per day, depending on the scenario.  

We can now estimate the range of released structures that are consistent with the observations. The horizontal dashed lines in Fig.~\ref{fig:fs1} give the observed range of four or five tracks per day. The resulting numbers  of released structures for each of the four  scenarios are compiled in Table~\ref{tab:table2}. The differences among the four scenarios are small. Scenarios 3 and 4, whose use is based on randomly drawn distributions for  input parameters of the simulations besides their speed, result in the wider ranges.

\begin{table*}[htp]
\begin{center}
\caption{Daily number of released structures in four simulated environments that are consistent with the number of detections in WISPR J-maps in Paper~1.}
\begin{tabular}{cc}
\hline 
\textbf{scenario} & \textbf{range} \\
1: periodic release longitudes and release times and  random speeds  & 44-113  \\
2: corotating release longitudes, periodic release times and random speeds & 42-96  \\
3: random release longitudes, periodic release times and random speeds & 37-119  \\ 
4: random release longitudes, release times and speeds & 35-127  \\ 
\hline
\label{tab:table2}
\end{tabular}
\end{center}
\end{table*}

Figures~\ref{fig:ftracks1} and \ref{fig:fs1} indicate that the most important factor in determining the number of track detections is the number of released density structures.
The dependence on the spatio-temporal release periods is much weaker since multiple  periods across extended ranges
correspond to the same number of detections.

A salient assumption in our analysis is that each track corresponds to a single density structure.
This translates to a lower limit on the number of released density structures that are consistent with the observations. This conclusion is supported by the reasonable expectation that the properties of some structures (size or intensity) \citep[see e.g. the simulations of][]{higginson2018} may drive them below  WISPR's detection thresholds in terms of spatial resolution, cadence, and/or sensitivity. We note, however, that  a close comparison of the WISPR time series with the associated J-map shows that there is a one-to-one correspondence
between density structures  in the movie and the corresponding J-map tracks.

\section{Discussion}

The weak correspondence between the number of detected tracks and both the considered scenario and the spatio-temporal release periods is unsurprising.We use the number of tracks to compare the synthetic and observed J-maps, which is expected to depend primarily on the number of released density structures rather than on the specifics of their release. This is clearly seen in Fig.~\ref{fig:fs1}, which shows a clear dependence of the number of detected tracks  on the number of released density structures. 

However, at relatively high daily numbers of released density structures ($> \approx$ 200), the number of detected synthetic tracks  becomes insensitive to that parameter. A factor influencing this could be the observing image cadence. In Fig.~\ref{fig:fcad} we display two synthetic J-maps produced by Scenario 1 where we consider the same speed (200 km/s) for all density structures. The only difference between the two J-maps is the cadence of 20 min (upper panel) and 1 min
(lower panel). The total number of released structures is $\approx$ 220 for both simulations and therefore  lies in the `saturated' portion of 
the number of released structures ($N_{ds}$)--number of detected tracks
($N_{tracks}$) curves in Fig. \ref{fig:fs1}. Clearly, the high-cadence J-map recovers more structures, something to
be investigated with higher-cadence observations. 

\begin{figure}[h]
\centering
\includegraphics[width=0.5\textwidth]{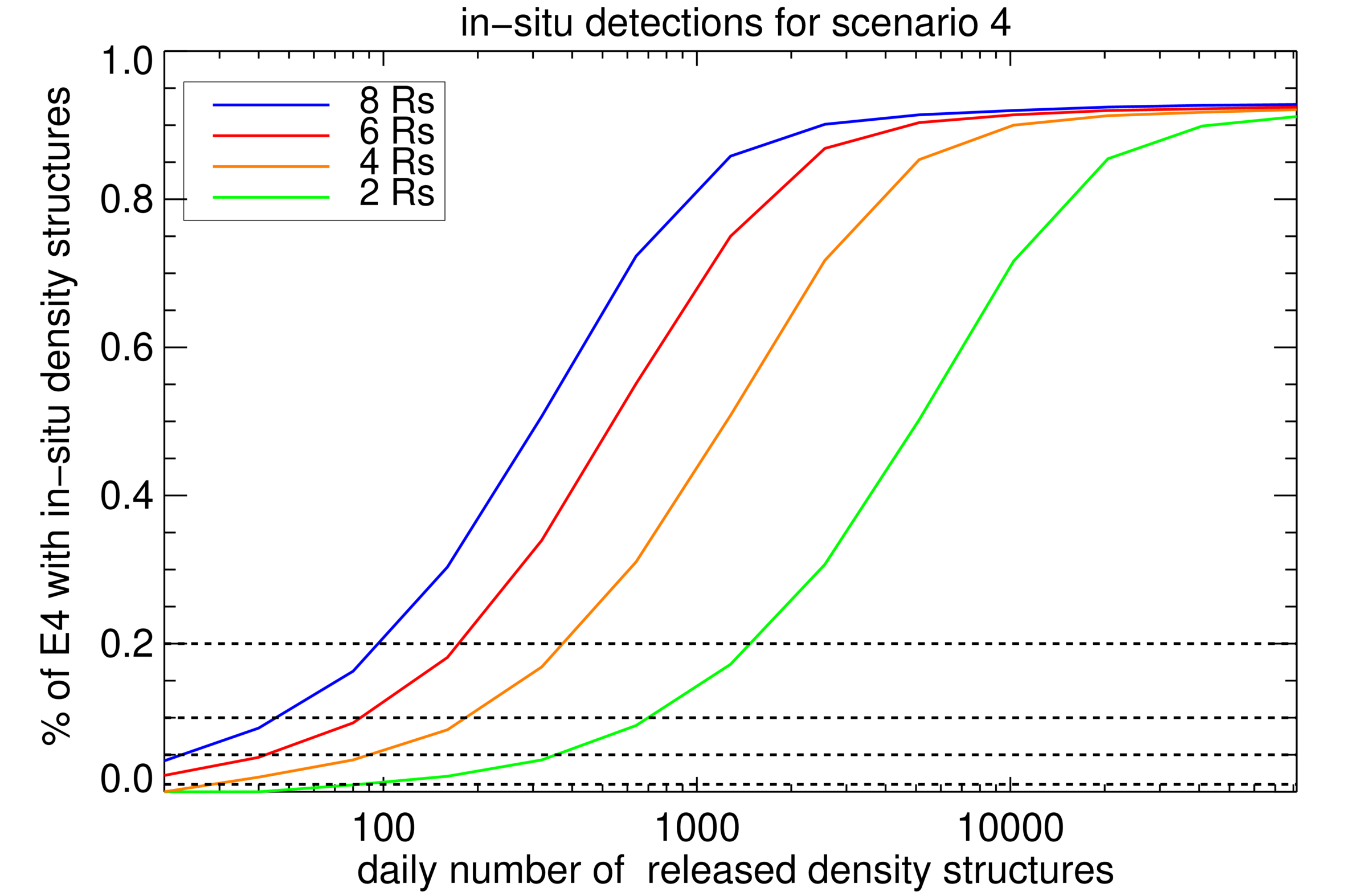}
\caption{50 \% quartiles 
of the percentage of the E4 duration
consistent with in situ detection
of density structures with radial sizes,
2 \rs (green line), 4 \rs (orange line), 6 \rs (red line),
and 8 \rs (blue line) as a function of the daily
number of released density structures from the
equatorial plane
for the 1000 Monte Carlo simulations for Scenario 4. The 
horizontal dashed lines (from bottom to top) correspond to 1, 5, 10, and 20 \% of the E4 duration.}
\label{fis}
\end{figure}

Our automated track detection scheme is based on the inferred value of temporal width based on the observations of J-map tracks from Paper 1.
However, we wonder how sensitive   our analysis is to the choice of this parameter.
In Fig. \ref{fig:fsen} we display the 50\% quartiles of the $N_{tracks}$
distributions 
from the  1000 Monte Carlo simulations of Scenario 4
as a function of $N_{ds}$, for four different choices 
of the J-map track duration: 60, 120, 240, and 480 minutes. We recall that
in our analysis we used a J-map track duration of 120 minutes (green line in Fig. \ref{fig:fsen}).
As expected, the shorter the considered track duration, the larger the number
of the detected tracks. For instance, the shorter
considered track duration (blue line) gives rise to a maximum
$N_{tracks}$  equal to 7;  conversely, the shorter considered
track width (brown line)  gives rise to maximum $N_{tracks}$  below 4, which is outside the range of the observations in Paper 1.
However, the intersections of the 
horizontal dashed lines at $N_{tracks}$ 4 and 5 with
the blue and green lines are consistent with rather small differences
of less than a factor of two in 
$N_{ds}$. The same applies to when no smoothing is applied
to the J-map and the 
same temporal width as in our original study is used (purple curve in Fig. \ref{fig:fsen}).
Taking all these together, this suggests the robustness of our approach. 
In retrospective, for  periods with faster and smaller density structures (i.e. when solar activity is higher than at E4) our framework could also be applied.
Finally, it is possible that
the plateaus of the $N_{tot}-N_{tracks}$ curves at large values of $N_{tot}$ discussed above could   also result from the 
line-of-sight
 overlap of multiple density structures. 
 
The derived number of released density structures
from the solar equatorial plane 
is about 35-130, which is clearly above
the estimates of \citet{sanchez2017}. 
These authors
used observations taken by  HI-1/SECCHI on board the  STEREO A spacecraft, corresponding to a period of highly inclined
heliospheric neutral lines during solar maximum conditions, and therefore enabling  a face-on
view of the streamer belt.
Using a combination of different maps,
\citet{sanchez2017} found that patches of enhanced emission at 30 \rs, which they
attributed to plasma blobs, occurred  
every 19.5 hours 
and were separated by 15 degrees in latitude. These findings suggested the release of  24 density structures 
from the streamer belt. 
One reason for the difference in the number of released
structures  may be linked
to the fact that a corotating interaction region (CIR)
was observed during the \citet{sanchez2017} observations.
However,   a possibly  much more important factor  is that
the \citet{sanchez2017} results are based on 
emission patches, which, 
as they also discuss, 
could be the amalgamation
of several individual smaller density structures.  The size of
the density structures of the
\citet{sanchez2017} study is $\approx$
12 $\times$ 5 \rs, whereas the density
structures of our WISPR observations
could be significantly smaller; for example,  the density
structure indicated with an arrow in panel (b) in Fig. 8 of Paper 1 is $\approx$
8 degrees in elongation and 4 degrees in
latitude (which  corresponds to $\approx$
5.6 $\times$ 2.8 \rs), as viewed from a PSP
viewpoint of 40 \rs. 

We now consider the implications from our J-map simulations for PSP in situ measurements of large-scale density structures during E4. We assume that a simulated density structure  would be detected in situ by PSP if its distance from PSP is comparable to the size of the density structure. 
Given that remote sensing observations are generally biased towards large-scale density structures, we considered density structure--PSP distances in the range of 2-8 \rs, consistent with inferred spatial scales of density structures from the WISPR-I E4 observations, as discussed above. By counting the number of timestamps leading to detections, we were able to estimate the percentage of the E4 
orbit that would contain in situ structure
detections. We note here that there could be timestamps populated by multiple density structures.

Our estimates are shown in  Fig. \ref{fis}, where the percentage of the E4 orbit with in situ structure detections is plotted against the number of released density structures in the solar equatorial plane for Scenario 4. The coloured  lines correspond 
to the 50 \% quartiles of the distributions resulting from the 1000  Monte Carlo simulations for in situ density structures--PSP distance (i.e. density structure size) of 2, 4, 6, and 8 \rs. The horizontal dashed lines correspond to   1, 5, 10, and  20  \% of the E4 orbit. The inferred daily number of released density structures
for Scenario 4 from the analysis of the J-maps in the previous section (i.e. 35-127; see Fig. \ref{fig:fs1} and Table 1)
  correspond to in situ detections for 1-20 \%  of the E4 duration. This prediction requires further investigation via a detailed analysis of the PSP E4 in situ density observations in a manner similar to past studies \citep[e.g.][]{viall2008,stansby2018} for observations at 1 au and 0.3-0.5 au, respectively.

\section{Summary}
Determining the global properties of transient slow solar wind density structures
is a key goal in the study of the slow solar wind. To this end, we performed simple Monte Carlo simulations for different scenarios regarding the properties of the 
released density structures, and compared the resulting synthetic J-maps against an observed J-map during PSP's fourth perihelion. Our findings can be summarized as follows:

\begin{enumerate}
\item The synthetic J-maps exhibit quasi-linear tracks similar to the observed tracks.
\item For scenarios employing periodic spatio-temporal density structure release, multiple spatio-temporal period pairs give rise to the same daily number of synthetic tracks as seen in the observations. 
The periods span extended intervals (i.e. 3\degr-45\degr\ and 1-25 hours). 
\item The four considered scenarios are consistent with similar ranges 35-45 to 96-127 of released density structures per day 
from the solar equatorial plane
(see Table 1) and  result in the same number of daily detected tracks as the observations in Paper~1.
For relatively high release rates ($> \approx$ 200), the number of detected synthetic tracks becomes insensitive to the release rate, and hence  the exact number of released density structures cannot be inferred from  WISRP-I E4 observations.
\item Our results predict that PSP in situ detections of density structures, with sizes in the range 2-8 \rs, will occur over 1-20 \% of the E4 orbit.
 \end{enumerate}

The estimates of the global release rates of density structures of this paper represent the first step in assessing their potential contribution to the slow solar wind mass budget. This will be achieved by calculations of the  mass content of the density structures observed by WISPR, which will be confronted with the corresponding in situ observations, and by further J-map calculations for other PSP as well as Solar Orbiter \citep[][]{muller2013} perihelia, in the latter scenario using EUI \citet{eui}, METIS \citep{Antonucci20}, and SolO-HI \citep{solohi} observations. Finally, it is important to compare PSP remote-sensing and in situ measurements of transient density structures to investigate whether the predictions of this work are correct.

\begin{acknowledgements}
The authors would like to thank the referee for useful comments and suggestions.
AV is supported by WISPR Phase-E funding to APL.We acknowledge the work of the PSP operations team. Parker Solar Probe was designed, built, and is now operated by the Johns Hopkins Applied Physics Laboratory as part of NASA’s Living with a Star (LWS) program (contract NNN06AA01C). We also acknowledge the efforts of the WISPR team developing and operating the instrument. SP and AV would like to thank AV for the invitation and hospitality during their visit to APL when this work was initiated. SP and AN acknowledge support by the ERC Synergy Grant 'Whole Sun' (GAN: 810218). 
\end{acknowledgements}

\bibliographystyle{aa}
\bibliography{jmap1}

\end{document}